\documentclass[aps,showpacs,twocolumn]{revtex4}
\usepackage{amsmath}
\usepackage{amsfonts}
\usepackage{amssymb}
\usepackage{graphicx}
\usepackage{indentfirst}
\usepackage{verbatim}

\begin{document}
\title{Controlling double ionization of atoms in intense bichromatic laser pulses}

\author{A. Kamor$^1$, F. Mauger$^{2}$, C. Chandre$^2$, and T. Uzer$^1$}
\affiliation{$^1$ School of Physics, Georgia Institute of Technology, Atlanta, GA 30332-0430, USA\\
$^2$ Centre de Physique Th\'eorique, CNRS -- Aix-Marseille Universit\'e, Campus de Luminy, case 907, 13288 Marseille cedex 09, France}
\date{\today}

\begin{abstract}
We consider the classical dynamics of a two-electron system subjected to an intense bichromatic linearly polarized laser pulse. By varying the parameters of the field, such as the phase lag and the relative amplitude between the two colors of the field, we observe several trends from the statistical analysis of a large ensemble of trajectories initially in the ground state energy of the helium atom: High sensitivity of the sequential double ionization component, low sensitivity of the intensities where nonsequential double ionization occurs while the corresponding yields can vary drastically. All these trends hold irrespective of which parameter is varied: the phase lag or the relative amplitude. We rationalize these observations by an analysis of the phase space structures which drive the dynamics of this system and determine the extent of double ionization. These trends turn out to be mainly regulated by the dynamics of the inner electron. 
\end{abstract}
\maketitle

\section{Introduction} \label{sec:Introduction}

Multiple ionization of atoms takes place at intermediate and high intensities when they are subjected to laser pulses~\cite{Beck08}. The routes by which these multiple ionizations occur can be quite complicated. Two main routes have been identified for the double ionization of two-active-electron systems (like the helium atom subjected to linearly polarized laser pulses): a sequential process and a nonsequential one~\cite{Beck08}. In the sequential process the field ionizes one electron after the other in an uncorrelated way. In contrast, the electron-electron correlation plays an active role in the ionization of the second electron in the nonsequential process. More specifically, the field picks up an electron, and sends this electron to the remaining ion core to dislodge the second electron, according the so-called recollision (or 3-step) scenario~\cite{Cork93, Scha93}. 

It has been observed experimentally in a broad range of intensities that non-sequential double ionization (NSDI) [also referred to as correlated double ionization (CDI)~\cite{Maug10}] yields can be many orders of magnitude larger than in the sequential double ionization (SDI) [also referred to as uncorrelated double ionization (UDI)] counterpart. Varying the intensity gives rise to a characteristic ``knee'' shape for the double ionization yield as a function of the laser intensity. This strong effect due to electron-electron correlation has been observed experimentally~\cite{Fitt92, Walk94, Webe00, Kond93, Laro98, Corn00, Guo01, DeWi01, Ruda04} and also theoretically using quantum~\cite{Beck08,Beck96, Beck99, Wats97, Lapp98, Panf03, Lein00, Feue01}, semi-classical~\cite{Fu01, Chen03, Barb96} and classical simulations~\cite{Sach01, Fu01, Ho05_1, Ho05_2, Panf03, Panf02, Liu07, Maug09, Maug10}. 

In this article, we investigate the dynamics resulting from the interaction between helium atoms and a bichromatic linearly polarized laser field using a classical model. Classical means are entirely sufficient to produce the strong electron correlation needed to explain double ionization phenomena~\cite{Ho05_1}.  Atomic phenomena in intense bichromatic fields can be considered as an extension of the principles of coherent control into the nonlinear nonperturbative regime of laser-atom interactions~\cite{Ehlo01, Shap03, Ko99, Howa91, Haff94, Petr00, Ivan95, Sirk01, Sirk02, Bati02, Koch03, Rang04, Cons05}. Although the classical mechanisms of double ionization remain qualitatively the same as for a monochromatic laser field~\cite{Maug09}, the double ionization yields show a strong sensitivity to changes in the parameters of the laser field, namely the phase lag between the two colors and their relative amplitudes. We explain this dependence using an analysis in terms of phase space structures. Our main result is that all these processes are regulated primarily by the dynamics of the inner electron which is the electron that remains bounded after the other one has been ionized. Remarkably, this electron, rather than being a spectator, is in fact the driver in the ionization process.

We consider a classical model with one spatial dimension for each electron which corresponds to the direction of polarization of the laser field. We denote the positions of the two electrons by $x$ and $y$ and set the origin at the nucleus. We call the canonically conjugate momenta of $x$ and $y$, $p_x$ and $p_y$ respectively. We consider soft Coulomb potentials~\cite{Java88,Ho05_1,Ho05_2,Panf02,Maug09, Maug10} defined by two softening parameters $a$ and $b$ for the interaction between charged particles. The Hamiltonian of the system in the dipole approximation is obtained by the sum of the kinetic energy of the electrons, the soft Coulomb potentials between the two electrons and between the electrons and the nucleus, and the interaction of the electrons with the field:
\begin{eqnarray}
   && H\left(x,y,p_x,p_y,t\right) = \frac{p_x^2}{2} + \frac{p_y^2}{2} + \frac{1}{\sqrt{\left(x-y\right)^2 + b^2}} \nonumber \\
   && \quad -\frac{2}{\sqrt{x^2 + a^2}} - \frac{2}{\sqrt{y^2 + a^2}} + \left(x+y\right) E\left(\omega t\right), \label{eq:Hamiltonian}
\end{eqnarray}
where $E\left(\omega t\right)$ is a bichromatic laser field, defined as:
\begin{equation} \label{eq:laserTerm}
   E\left(\omega t\right) = E_0 f(\omega t) \left(\sin\left(\omega t\right) + \alpha \sin\left(k \omega t + \delta \right) \right),
\end{equation}
where $E_0$ is the amplitude of the dominant mode, $\omega$ its frequency, $\alpha<1$ the ratio of the amplitudes of the two colors of the laser field, $k$ characterizes the mode locking between the two components, and $\delta$ is the phase lag. Since the second laser is an additional energy input to the system, the intensity is scaled with $\alpha$ as $I_{0} \propto E_{0}^{2}\left(1+\alpha^{2}\right)$. The system has two and a half degrees of freedom, one for each electron and one half for time dependence. The envelope function $f$ is chosen as a trapezoid with 2 laser cycle (l.c.) ramp up, 4 laser cycle plateau, 2 laser cycle ramp down, and zero elsewhere. In what follows, the wavelength is chosen to be 780 nm (corresponding to a frequency of $\omega=0.0584$ a.u.) and $k=3$. Except when we investigate the role of $\alpha$, we set $\alpha = 0.45$ for numerical simulations, which corresponds to an additional intensity input of 20\%. We consider the initial state to be the ground state of He, i.e. ${\cal E}_g=-2.24$ a.u.~\cite{Haan94, Ho05_1}. We choose $a=1$ to prevent auto-ionization (so that the energy surface is bounded in phase space). The second softening parameter $b$ is chosen so as to allow a significant energy exchange during a recollision. Here $b=1$~\cite{Panf01}. 

For insight into the dynamics of typical trajectories and to interpret the mechanisms at play which explain the statistical analysis, we present two kinds of studies as functions of the parameters of the field, $E_0$, $\alpha$ and $\delta$ in what follows: In Sec.~\ref{sec:stat}, a statistical analysis of a large ensemble of trajectories, and in Sec.~\ref{sec:NlDynAnalysis} an analysis of the phase space structures.

\section{Statistical analysis} \label{sec:stat}

Starting from the classical model~(\ref{eq:Hamiltonian}), we perform a statistical analysis of a large ensemble of trajectories (typically 150,000) initially on the ground state (with a microcanonical distribution~\cite{Maug09}). At the end of the pulse, a trajectory can exhibit zero, one or two ionized electrons. In our calculations, an electron is considered ionized when its energy (defined as the sum of its kinetic energy and the Coulomb interaction with the nucleus) is positive, which corresponds to an electron free from the Coulomb potential well. A convenient observable is the probability of double ionization where the two electrons have a positive energy at the end of the pulse. This probability is commonly plotted as a function of the laser intensity. A signature of NSDI is a characteristic knee shape: In the intermediate intensity range, these yields are much higher (by several orders of magnitude) than expected from the uncorrelated sequential double ionization mechanism~\cite{Beck08}. At high intensity, the yields reach saturation, corresponding to double ionization of all the electrons of the ensemble. In what follows, we investigate the double ionization yields versus the intensity when the phase lag $\delta$ and the relative amplitude $\alpha$ are varied.

\subsection{Varying the phase lag $\delta$} \label{sec:vary_delta}

\begin{figure}
   \centering
   \includegraphics[width = \linewidth]{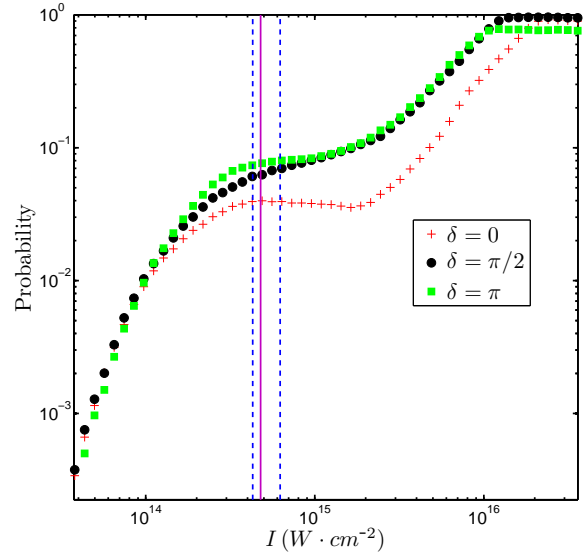}
	 \caption{\label{fig:knee_delta} 
	 Probability of double ionization as a function of the intensity for $\alpha=0.45$.	The red crosses correspond to $\delta = 0$, the black circles to $\delta=\pi/2$, the green squares to $\delta=\pi$. We also indicate the range of variation of $I^{(c)}$ with $\delta$ as given by Eq.~(\ref{eq:EqualSharingRelation}) (vertical dashed-dotted lines). The vertical magenta continuous line labels the intensity chosen in Fig.~\ref{fig:Max_P_DI_delta},~\ref{fig:Laminar}, and~\ref{fig:action_delta}.}
\end{figure}

In Fig.~\ref{fig:knee_delta} we display several double ionization probability curves as a function of the laser intensity for different values of the phase lag $\delta$ (for $\alpha=0.45$).

In the intermediate range of intensities (up to saturation of double ionization), we observe a significant variability of the double ionization probabilities and consequently of the characteristic features of the knee. More precisely, in the intermediate intensity range the level of double ionization yields greatly vary with $\delta$ by a factor of 3; however, when  varying $\delta$ the range of intensity where the knee is located (around $5\times 10^{14} \ {\rm W}\cdot{\rm cm}^{-2}$, vertical dashed-dotted lines in Fig.~\ref{fig:knee_delta}) does not change significantly. In the low intensity regime i.e., for intensities lower than $8 \times 10^{13} \ {\rm W} \cdot {\rm cm}^{-2}$ there is a very weak dependence of $\delta$ in these yields. The same conclusions hold for other values of $\alpha$. We note a significant impact of the phase lag in the resulting double ionization probability, meaning that adjusting this parameter can act as a control knob for both sequential and non-sequential processes. One advantage of varying the phase lag between the two components of the laser field is that no additional input of energy is required (as opposed to varying other parameters like $\alpha$) since the mean value of the electromagnetic energy does not depend on $\delta$.    

\begin{figure}
   \centering
   \includegraphics[width = \linewidth]{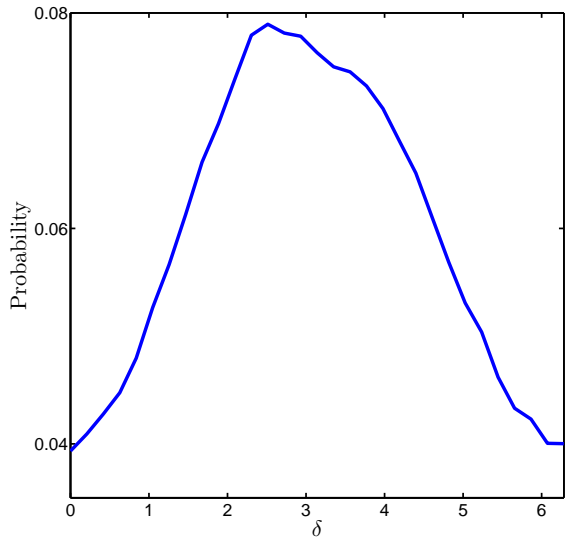}
   \caption{ \label{fig:Max_P_DI_delta}
   Double ionization probability as a function of $\delta$ for $\alpha = 0.45$ and the intensity labeled by the vertical magenta continuous line in Fig.~\ref{fig:knee_delta} ($I = 4.78 \times 10^{14} \ {\rm W}\cdot{\rm cm}^{-2}$).}
\end{figure}

In Fig.~\ref{fig:Max_P_DI_delta} we represent the double ionization yield as a function of $\delta$ for the intensity labeled by the vertical magenta continuous line in Fig.~\ref{fig:knee_delta}. The resulting curve has a bell shape with a maximum slightly shifted before $\delta=\pi$. More specifically, we observe a fast variation of the double ionization yield with changes of the phase lag around $\delta=0$ and a flattening around $\delta=\pi$. This confirms the high sensitivity of the double ionization probability to the parameter $\delta$. 

Except for very low intensity (i.e. below $8\times10^{13}\ {\rm W}\cdot{\rm cm}^{-2}$) we observe qualitatively the same impact of $\delta$ on the double ionization yield (a bell shape almost centered around $\delta=\pi$). The precise value of the phase lag for which the probability is maximum depends on the laser parameters.

The value of $\delta$ for which we observe the first saturation of double ionization occurs when $\delta=\pi$. This value of $\delta$ does not depend on $\alpha$ (and is defined as soon as $\alpha$ is non-zero). We define $I^{\rm (t)}\left(\delta\right)$ the intensity, as a function of the phase lag, for which saturation of double ionization is reached. Then, at high intensity, we note a great variability of $I^{\rm (t)}\left(\delta\right)$ (the differences are of the order of $10^{16}\ {\rm W}\cdot{\rm cm}^{-2}$) which is more than one order of magnitude larger that the range of variation of the knee.

\subsection{Varying the relative amplitude $\alpha$}

\begin{figure}
   \centering
   \includegraphics[width = \linewidth]{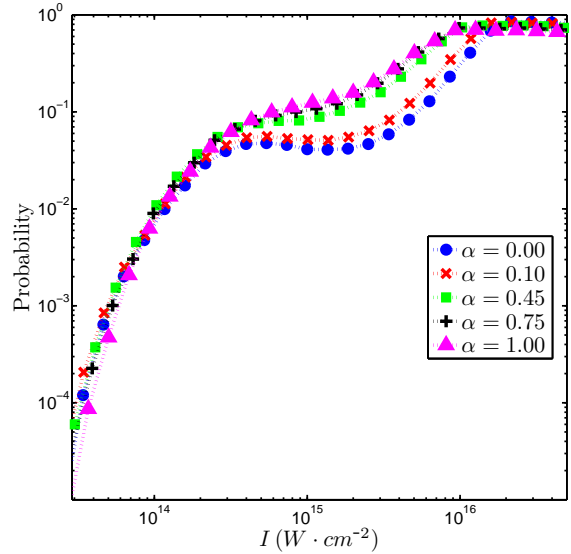}
	 \caption{ \label{fig:local_max_CDI}
	 Double ionization probability as a function of intensity for different values of $\alpha$. Blue circles correspond to~$\alpha=0$ (single laser), Red crosses to~$\alpha=0.10$, green squares to~$\alpha=0.45$, black plus signs to~$\alpha=0.75$ and magenta triangles correspond to~$\alpha=1.00$. Here $\delta=\pi$ for all curves.}
\end{figure}

\begin{figure}
   \centering
   \includegraphics[width = \linewidth]{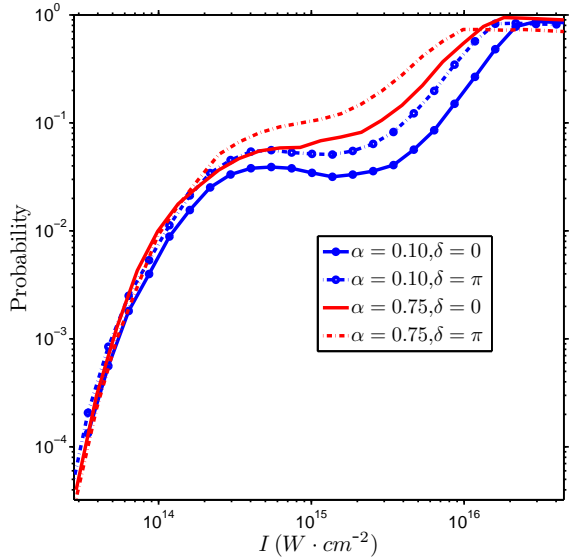}
   \caption{\label{fig:alphaControl}
   Double ionization probability as a function of the intensity for different values of $\delta$. We keep $\alpha$ constant and equal to $\alpha=0.1$ and $\alpha=0.75$, respectively.}
\end{figure}

After varying $\delta$ at fixed $\alpha$, we keep $\delta$ fixed at $\pi$ in order to investigate $\alpha$'s influence on the double ionization probability. We chose $\delta = \pi$ because it corresponds to the phase lag where the peak field amplitude is maximum (see Sec.~\ref{sec:DI_proba}). In Fig.~\ref{fig:local_max_CDI} we represent the double ionization probability as a function of the intensity for different values of $\alpha$. As for the variation of $\delta$, we notice no significant dependence on the parameters at low intensities (below $8\times 10^{13}\ {\rm W} \cdot {\rm cm}^{-2}$). In the intermediate intensity range, we notice that the probability greatly varies with $\alpha$ (more than a factor two) whereas the intensity at which the characteristic knee is developed does not change significantly. In the meantime, the UDI component also increases with $\alpha$, and we notice a decrease of the intensity $I^{\rm (t)}$ where double ionization is saturated. As a consequence we notice a flattening in the knee, the two CDI and UDI curves merging. 
As $\alpha$ is increased, the second color of the laser field becomes more important, hence $\delta$ is expected to have a more prominent role in controlling the system because the variation in probability becomes larger. This is confirmed by Fig.~\ref{fig:alphaControl} where we have represented double ionization probabilities as a function of the intensities for $\alpha=0.1$ and $\alpha=0.75$.

In summary, the statistical analysis shows these three main trends when changing the parameters of the field:
\begin{enumerate}
   \item High sensitivity of the actual probability of double ionization, \label{Obs:3}
   \item High sensitivity of the intensity of saturation of the double ionization yields, \label{Obs:1}
   \item Low sensitivity of the intensity of knee regime (i.e. the CDI maximum). \label{Obs:2}
\end{enumerate}
In the next section, we analyze the dynamics of Hamiltonian~(\ref{eq:Hamiltonian}) in terms of phase space structures and we find results that support and explain these trends.

\section{Nonlinear dynamical analysis}\label{sec:NlDynAnalysis}

For the dynamical analysis of Hamiltonian~(\ref{eq:Hamiltonian}) we follow the one performed in Ref.~\cite{Maug09} for the monochromatic laser pulse. The analysis proceeds in the following way: First, without laser field ($E_{0}=0$), an outer electron and an inner one were identified from a periodic orbit analysis of the chaotic motion of typical trajectories. When the field is turned on, if the inner electron remains bounded, the atom may undergo a recollision which follows from the 3-step picture~\cite{Cork93, Scha93} (the field picks up the outer electron, and upon reversal of the field, brings it back to the core and the two electrons share the collision energy).

\subsection{Effective models}

The dynamics of both electrons are accurately reproduced by reduced models except during recollisions~\cite{Maug09}. These models allow one to derive the main properties of the double ionization curves versus intensity. We begin by investigating the behavior of typical trajectories. In Fig.~\ref{fig:CDI_Trajectory}, we represent the position of each electron (upper panel) and their energy (kinetic energy plus soft Coulomb potential, lower panel) as functions of time for a typical trajectory. We observe that one electron is first quickly ionized (the ``outer'' electron) while the other remains close to the core (the ``inner'' electron). Once ionized, the outer electron is driven mainly by the laser field and experiences large excursions from the core. In contrast, the inner electron experiences a competition between Coulomb interaction with the nucleus and the laser excitation. Without loss of generality, we assign $x$ to the outer electron and $y$ to the inner one. In the sample trajectory of Fig.~\ref{fig:CDI_Trajectory} the inner electron (red curves) does not ionize until it collides with the outer electron (blue curves).  Double ionization occurs at approximately 2.5 laser cycles, which can be confirmed by examining the energies of both electrons in the lower panel.  We see that both electrons have positive energy after this time (and thus are ionized).
From this qualitative analysis, two reduced effective Hamiltonians are identified, $H_{\rm out}$ for the outer and $H_{\rm in}$ for the inner one:   
\begin{figure} 
   \centering
   \includegraphics[width = \linewidth]{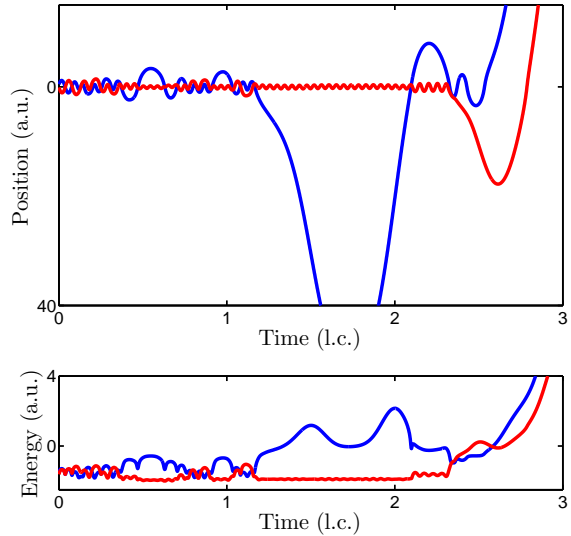}
	\caption{ \label{fig:CDI_Trajectory}
	Typical nonsequential double ionization. We display the position (upper panel) and energy (as defined in the text, lower panel) of each electron as functions of time.  The parameters are $I = 4.78 \times 10^{14} \ {\rm W} \cdot {\rm cm}^{-2}$, $\alpha = 0.45$ and $\delta = 0$.}
\end{figure}
\begin{eqnarray}
   H_{\rm out} \left(x,p_x,t\right) & = & \frac{p_x^2}{2} + x E\left( \omega t\right), \label{eq:H_out} \\
   H_{\rm in}  \left(y,p_y,t\right) & = & \frac{p_y^2}{2} - \frac{2}{\sqrt{y^2 + 1}} + y E\left(\omega t\right), \label{eq:H_in}
\end{eqnarray}  
where we have neglected the interaction with other charges for the outer electron and have allowed the inner electrons to feel both the laser field and the attraction from the nucleus. For both electrons, we ignore the electron-electron interaction term since they are assumed to be far away from each other between two recollisions. The main advantage of the above effective models is that their dimensions are lower than the one of Hamiltonian~(\ref{eq:Hamiltonian}) (they both have one and a half degrees of freedom). Hamiltonian $H_{\rm out}$ is even integrable in the sense that one can explicitly give the expression of the trajectories as functions of the initial conditions. Hamiltonian $H_{\rm in}$ is not integrable but its dynamics can be analyzed qualitatively. In what follows, we analyze these two models as the parameters of the field are varied and correlate their dynamics with the double ionization curves plotted in Sec.~\ref{sec:stat} in order to support and explain the observed trends.

\subsection{dynamics of the outer electron}
\begin{figure}
   \centering
   \includegraphics[width = \linewidth]{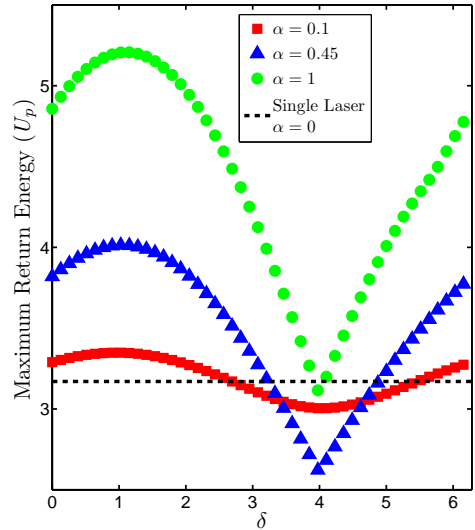}
   \caption{ \label{fig:max_return_outer_elec}
   Maximum return energy of the outer electron as a function of $\delta$ in units of $U_p$.  Red squares correspond to $\alpha=0.10$, blue triangles to $\alpha=0.45$, and green circles to $\alpha=1.00$, while the dashed black line corresponds to the one laser system ($\alpha=0$).}
\end{figure}

The trajectory of the outer electron [as governed by Eq.~(\ref{eq:H_out})] is given by
\begin{eqnarray}
   x\left(t\right)   & = & x_0 + p_0 t + \frac{E_0}{\omega^2}\left(\sin\omega t + \frac{\alpha}{k^2}\sin\left(k\omega t + \delta \right) \right), \\
   p_x\left(t\right) & = & p_0 + \frac{E_0}{\omega}\left(\cos\omega t + \frac{\alpha}{k} \cos \left( k \omega t + \delta \right) \right),
\end{eqnarray}
where $x_{0}+E_{0} \alpha k^{-2}\omega^{-2} \sin \delta$ and $p_{0} + E_0 \omega^{-1}\left( 1 + \alpha k^{-1} \cos \delta \right)$ are, respectively, the position and momentum of the outer electron at time $t=0$. An important quantity is the maximum return kinetic energy of the outer electron: It gives the maximum amount of energy the outer electron can bring to the inner electron to trigger CDI. In addition, if the outer electron recombines, it gives the cutoff in harmonic generation for photon emission~\cite{Band05}. In order to compute this quantity we need to consider a shift of the field since the electron leaves the core at a certain phase $\phi$ of the field. The effective Hamiltonian is transformed into
$$
   H_{\rm out}=\frac{p_x^2}{2}+x E(\omega t +\phi).
$$ 
Maximizing the kinetic energy ${\cal E}_{\rm max}=p_x(t_r)^2/2$ with respect to the recollision time $t_{r}$ and the initial phase $\phi$ leads to the familiar condition that the maximum return energy is associated with trajectories leaving the nucleus with zero momentum~\cite{Band05}. As a function of the parameters of the laser field, this maximum kinetic energy is equal to
\begin{equation}
    {\cal E}_{\rm max} = \kappa\left(\alpha,\delta\right) U_p,
\end{equation}
where $U_p=E_0^2/(4\omega^2)$ is the ponderomotive energy and $\kappa$ is only a function of $\alpha$ and $\delta$. For $\alpha=0$, we have $\kappa(0,\delta)\approx 3.17$~\cite{Cork93}. In Fig.~\ref{fig:max_return_outer_elec}, we represent $\kappa(\alpha,\delta)$ as a function of $\delta$ for various values of $\alpha$. We notice that the amplitude of the variations of $\kappa$ with $\delta$ increases with $\alpha$. For a fixed value of $\alpha$, the maximum return energy $\kappa$ (in units of $U_p$) of the outer electron can vary significantly with $\delta$ (up to more than 2). For all values of $\alpha$, the maximum of $\kappa$ occurs at $\delta \approx 1.04$, while its minimum is obtained for $\delta \approx 3.99$. We notice that none of these values are related to the critical values of the phase lag for which we observe maximum of double ionization (around $\delta=\pi$) and saturation of double ionization. Thus, the origin of the amount of double ionization does not solely rely on the dynamics of the outer electron. It has to be complemented by the dynamics of the inner electron.

\subsection{Dynamics of the inner electron}

\begin{figure*}
   \centering
   \includegraphics[width = .49 \linewidth]{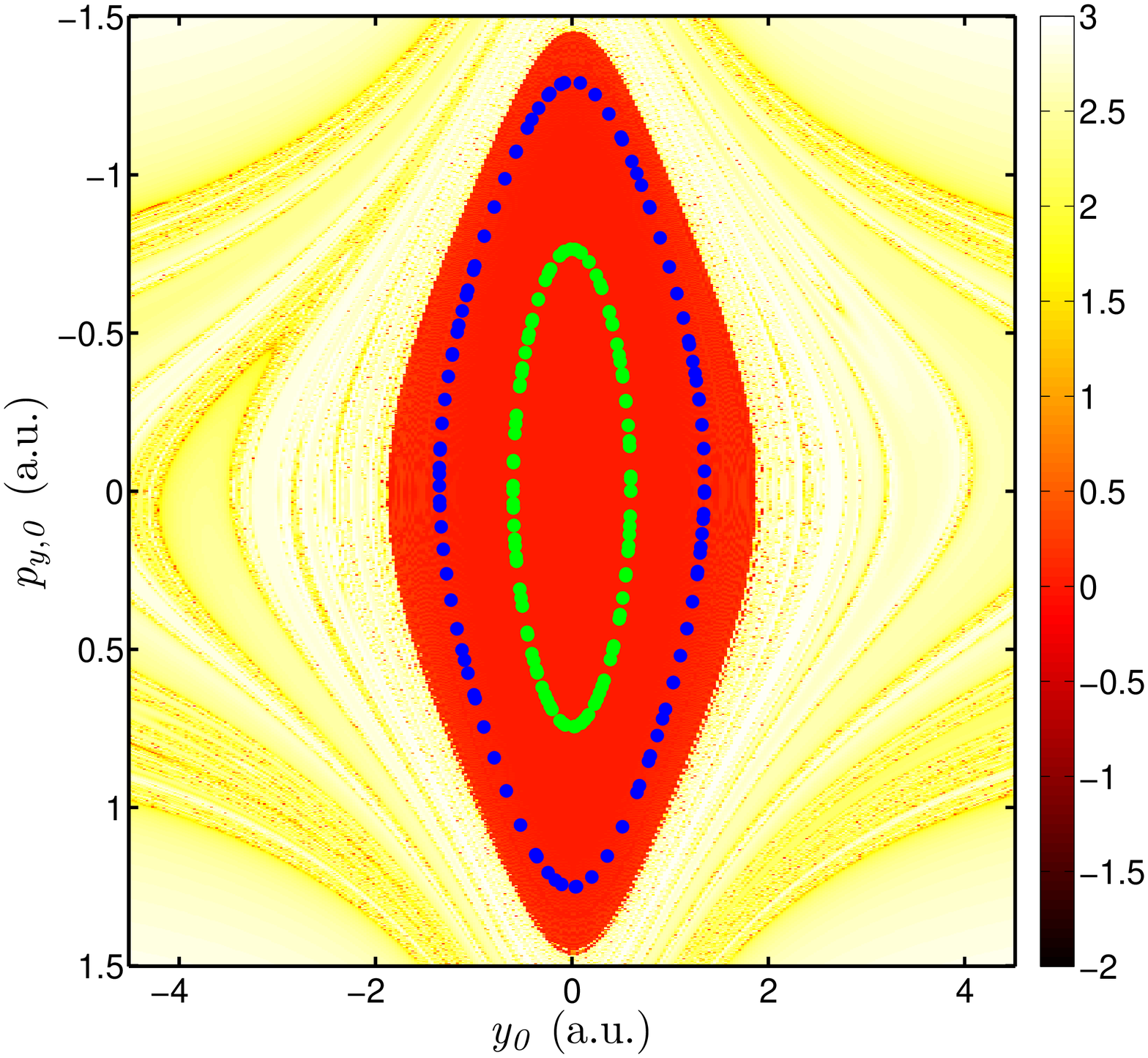}
   \includegraphics[width = .49 \linewidth]{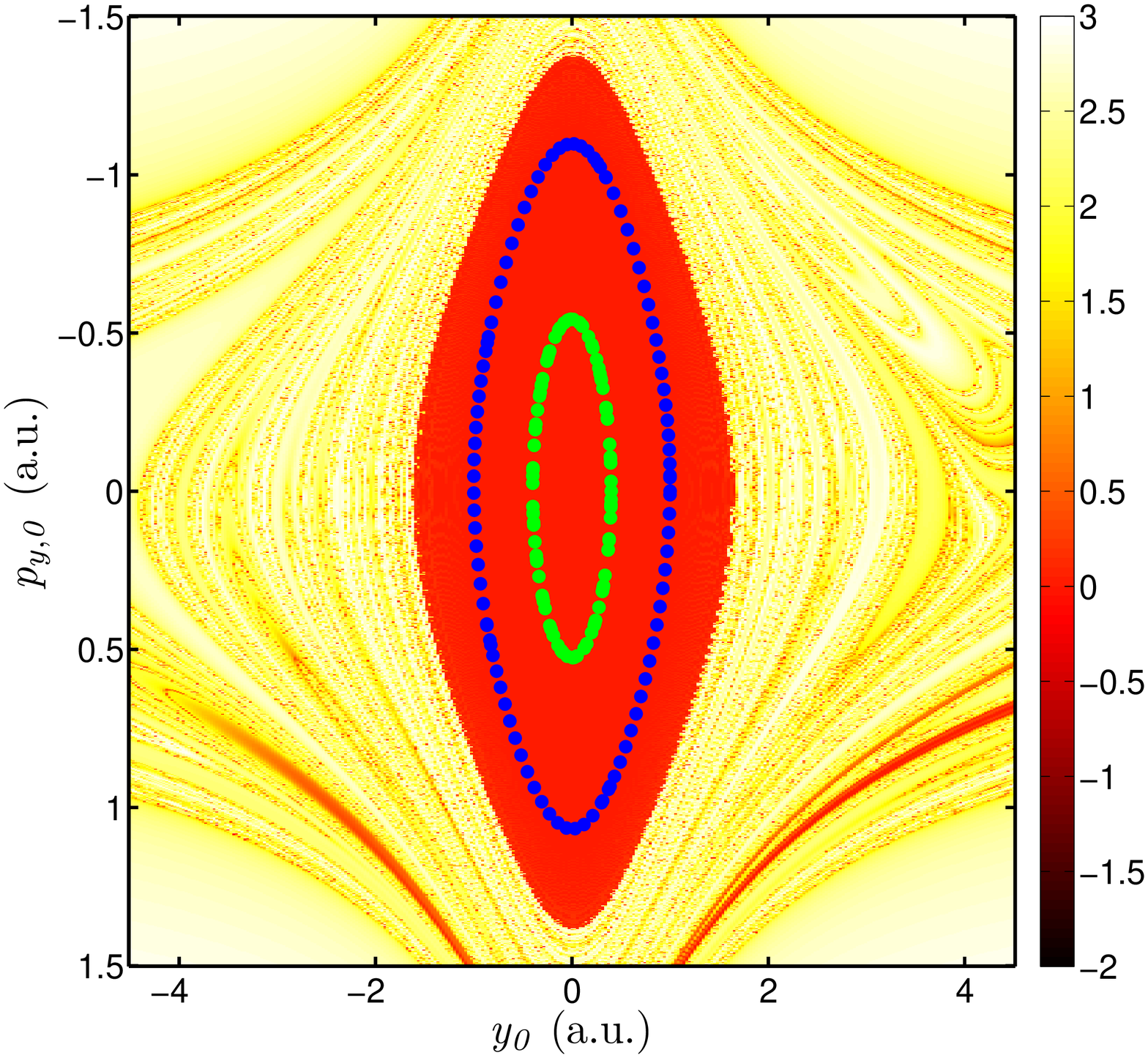}
   \caption{\label{fig:Laminar}
   Laminar plots of inner electron dynamics as given by Hamiltonian~(\ref{eq:H_in}) as a function of the initial conditions $(y_0,p_{y,0}, t=0)$.  For each initial condition, we represent the distance of the electron from the nucleus after 10 laser cycles.  The distance is plotted in a logarithmic scale.  In the left panel $\delta = 0$ and in the right panel $\delta = \pi$.  The dashed curves are Poincar\'e sections of two typical trajectories in the bound region taken stroboscopically with the laser period. The parameters are $\alpha=0.45$ and $I=4.78 \times 10^{14} \ {\rm W} \cdot {\rm cm}^{-2}$ in both plots.} 
\end{figure*}

After the ionization of the outer electron, the inner electron experiences the combined action of the field and of the attraction from the nucleus only. Its Hamiltonian is given by Eq.~(\ref{eq:H_in}). The competition between these two interactions leads to two distinct behaviors related to two distinct regions. Figure~\ref{fig:Laminar} represents these two regions in phase space: More specifically, we plot the distance of the inner electron from the nucleus after 10 laser cycles in the space of initial conditions $(y_{0},p_{y,0})$ at $t=0$. Note that the color map is in logarithmic scale. The red and green curves are two different Poincar\'e sections of an inner electron trajectory, taken stroboscopically with the laser period. They indicate that the dark region is filled with invariant tori where electrons are bound to the nucleus (for all time). The light region corresponds to ionizing electrons (after a transient time which depends on the initial conditions, see the darker structures which spiral out). The rationale behind this dynamics is as follows: If the electron is close to the nucleus, then the Coulomb interaction dominates over the field and the electron is bound. If it is sufficiently far, the field dominates and the dynamics of the electron is unbounded, in a similar way as the outer electron. The maximum distance the inner electron can be from the nucleus while remaining bound is a function of the laser parameters. 

\subsection{Double ionization probability} \label{sec:DI_proba}

\begin{figure}
   \centering 
   \includegraphics[width = \linewidth]{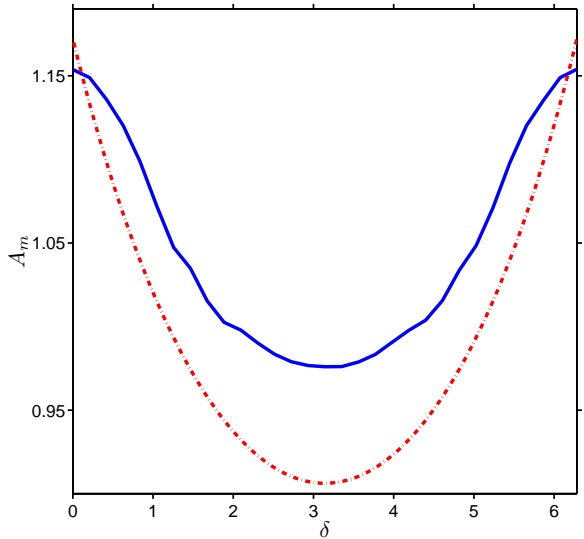}
   \caption{ \label{fig:action_delta} 
   Action $A_m$ of the outermost invariant torus as a function of $\delta$ for $\alpha=0.45$ and $I=4.78 \times 10^{14} \ {\rm W} \cdot {\rm cm}^{-2}$.  The solid line corresponds to the statistical data, while the dashed dotted line corresponds to the theoretical model, as given by Eq.~(\ref{eq:ThAm}).}
\end{figure}

Varying certain laser parameters leads to strong variations in the size of the bound region, which is pivotal in explaining the probability of double ionization. Hamiltonian~(\ref{eq:H_in}) is not integrable. However, a close inspection of the dynamics of the inner electron in the bound region reveals that it is fairly insensitive to the field so we consider another effective Hamiltonian in the bound region, namely
\begin{equation}\label{eq:FreeFieldInnerElectron}
   H^{*}_{{\rm in}}\left(y,p_{y}\right) = \frac{p_{y}^{2}}{2} - \frac{2}{\sqrt{y^{2} + 1}}, 
\end{equation}
which is integrable, its phase space being foliated by periodic orbits. There exists, at least locally, a canonical transformation which maps Hamiltonian~(\ref{eq:FreeFieldInnerElectron}) into action and angle variables. The action is defined by
\begin{equation*}
   A = \frac{1}{2 \pi} \oint{p_{y} dy}.
\end{equation*}
When the field is present, most of the outer invariant tori are broken, whereas most of the inner ones close to the nucleus persist. We denote by $A_{m}\left(E_{0},\alpha,\delta\right)$ the action of the last invariant rotational torus which roughly corresponds to the boundary of the inner region. Figure~\ref{fig:action_delta} illustrates the dependence of $A_m\left(E_{0},\alpha,\delta\right)$ with $\delta$ as $\alpha$ and $E_{0}$ are kept fixed for an intensity in the UDI regime ($I=4.78\times{10}^{15}\ {\rm W}\cdot{\rm cm}^{-2}$). Numerically, we  determine $A_{m}$ by computing the volume of phase space for which the inner electron never ionizes  during 2 laser cycle ramp up and 4 laser cycle plateau (see Fig.~\ref{fig:action_delta}). At least qualitatively, the same trend is observed in the whole UDI regime (as well as the CDI regime), namely an inverted bell shape with fast variations of $A_{m}$ around $\delta=0$ with a flattening and minimum centered around $\delta=\pi$.

\begin{figure}
   \centering
   \includegraphics[width = \linewidth]{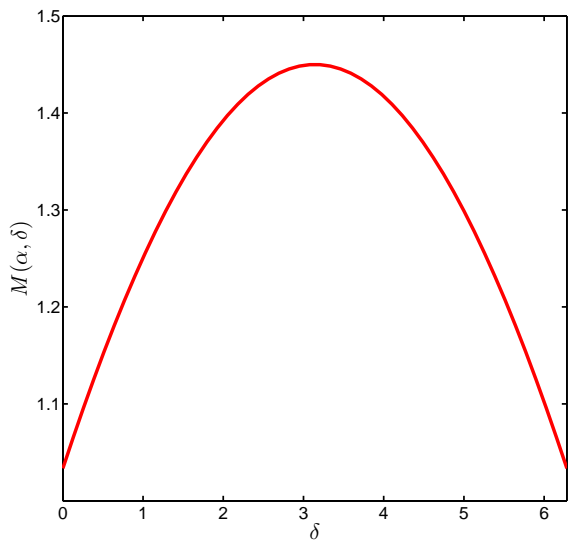}
   \caption{ \label{fig:Laser_maximum_delta} 
   $M\left(\alpha = 0.45, \delta\right)$ as a function of $\delta$, as given by Eq.~(\ref{Eq:M_alpha_delta}).}
\end{figure}

An approximation to the size $y_{m}$ of the inner region can be obtained by finding the local maximum of the effective potential (soft Coulomb potential plus laser excitation) at peak field amplitude. It is implicitly given by
\begin{equation} \label{eq:expansionApprox}
   M\left(\alpha, \delta\right) E_0 = \frac{2y_m}{\left(y_m^2+1\right)^{3/2}},
\end{equation}
where $M(\alpha, \delta)$ is given by:
\begin{equation} \label{Eq:M_alpha_delta}
   M\left(\alpha,\delta\right) = \max_{\tau\in\left[0,2\pi\right]}\left| \sin \tau+\alpha \sin\left(k\tau+\delta\right)\right|,
\end{equation}
and corresponds to the peak field amplitude of the laser. In Fig.~\ref{fig:Laser_maximum_delta} we display the peak field  amplitude as a function of $\delta$. 
In the bound region, the energy of the inner electron, as defined by Eq.~(\ref{eq:H_in}), does not vary much. Then, it is possible to relate it to the action $A$ through the formula~\cite{Maug10}
$$
   H_{\rm in} \left( A \right) = -2 + \sqrt{2} \frac{{\rm e}^{\gamma A} - 1}{\gamma},
$$
where $\gamma=-9\sqrt{2}/16$~($\approx-0.8$). It can be inverted to express the action explicitly as a function of the energy
\begin{equation} \label{eq:energy2action}
   A = \frac{1}{\gamma}\ln{\left(1 + \frac{\gamma}{\sqrt 2}\left(H_{\rm in} + 2\right)\right)}.
\end{equation}
The outermost position $y_{m}$ at peak field amplitude is associated with zero momentum, such that its energy $H_{m}$ is equal to
$$
   H_{m} = H_{\rm in}\left(y_{m}, 0, t_{m}\right) = -\frac{2}{\sqrt{y_{m}^{2}+1}} - E_{0} M\left(\alpha,\delta\right) y_{m}.
$$
Finally, the critical action $A_{m}$ is obtained by plugging $H_{m}$ into Eq.~(\ref{eq:energy2action})
\begin{equation}\label{eq:ThAm}
   A_{m} = \frac{1}{\gamma}\ln\left(1 + \frac{\gamma}{\sqrt{2}} \left( 2 - \frac{4y_{m}^{2}+2}{\left(y_{m}^{2}+1\right)^{3/2}}\right)\right),
\end{equation}
where $y_{m}$ is solution of Eq.~\ref{eq:expansionApprox}. We compare this prediction with numerical evaluation of the action in Fig.~\ref{fig:action_delta}. A careful analysis of $A_{m}$, as given by Eq.~(\ref{eq:ThAm}) shows that $A_{m}$ is symmetric and minimum around $\delta=\pi$ for all $\alpha$.

Sequential double ionization (UDI) corresponds to the situation when one electron is first ionized by the field, leaving the inner electron in the unbound region, so that the field sequentially ionizes the remaining electron. The amount of UDI is related to the volume of the bound region, therefore to $A_m$: the smaller the inner region (and thus $A_{m}$) the more likely it is for the inner electron to be found in the unbound region as well. Nonsequential (CDI) double ionization corresponds to the situation when the inner electron is first left in the inner region, and is then propelled, through a recollision, in the unbound region under the impact of the returning outer electron. As a result, it is easier for the outer electron to ionize the inner one if the bound region is small. Either way, for both CDI and UDI, we are expecting higher double ionization probability when $A_{m}$ is minimum (i.e. around $\delta=\pi$) and a strong sensitivity of the double ionization yield where $A_{m}$ varies fast (i.e. around $\delta = 0$). It corresponds to what is observed in Fig.~\ref{fig:Max_P_DI_delta}. The small discrepancy between $\delta = \pi$ and the actual maximum is due to nonlinear energy exchange processes during the recollision. In Fig.~\ref{fig:max_return_outer_elec}, we see that the kinetic energy of the outer electron upon return is not symmetric about $\delta=\pi$. In particular we see that the maximum return energy is rather small around $\pi$ in comparison with slightly lower values of $\delta$ which explains that the maximum is expected to be slightly smaller than $\pi$.

The rate at which $A_m$ changes with $\delta$ provides an explanation for the sensitivity of the probability to changes in $\delta$ which was discussed in Sec.~\ref{sec:vary_delta}.  The action changes quickly in $\delta$ in the regions around $\delta=0$ and slowly around $\delta=\pi$, which is the same trend observed for the rate of change in probability with $\delta$ from Fig.~\ref{fig:Max_P_DI_delta}. This explains Observation~\ref{Obs:3} from the statistical analysis, in Sec.~\ref{sec:stat}, namely that the double ionization yield is highly sensitive to the choice of phase lag.

\subsection{Saturation of the double ionization probability}

As explained before, CDI corresponds to the case where both electrons are left in the unbound region. An estimate of the size of the bound region is implicitly obtained through Eq.~(\ref{eq:expansionApprox}). An approximation of the amplitude of the field at which there is no inner region leads to an estimate of the intensity at which double ionization is saturated, meaning that all the trajectories double ionize. Equation~(\ref{eq:expansionApprox}) has a real solution if $E_{0} M$ is smaller than $4/(3\sqrt{3})$ which leads to
$$
E_0^{\rm (t)}(\alpha,\delta)=\frac{4}{3\sqrt{3}M(\alpha,\delta)}.
$$
The field is maximum for $\delta=\pi$ and its value is $M(\alpha,\pi)=1+\alpha$. For this value of $\delta$ the critical value of the intensity at which all the trajectories double ionize is minimum. This is what is observed in Fig.~\ref{fig:knee_delta} and~\ref{fig:alphaControl}. A numerical estimate of the range of variation of this critical intensity as a function of $\delta$, corresponding to the amplitude of variation of intensity between $E_0^{\rm (t)}(\alpha,0)$ and $E_0^{\rm (t)}(\alpha,\pi)$ gives $1.1\times 10^{16}\ {\rm W} \cdot {\rm cm}^{-2}$ at $\alpha =0.45$ which is what is observed in Fig.~\ref{fig:knee_delta}. A more precise approximation of the value of the intensity (as a function of the laser parameters) at which the bound region disappears is obtained by following the elliptic periodic orbit (with the period of the laser field) at the center of the elliptic island~\cite{Maug09}. It explains Observation~\ref{Obs:1} from the statistical analysis,  in Sec.~\ref{sec:stat}, namely that the intensity of saturated double ionization greatly varies with the  choice of the laser parameters.

\subsection{Maximum of CDI}

When one electron is bound to the nucleus after the other one has ionized, the only way through which it leaves the nucleus is by an exchange of energy with the outer electron when this one comes back to the core upon sign reversal of the laser field. Through this recollision the inner electron jumps from one invariant torus to another one (depending on the amount of energy exchanged), and it can even jump from one invariant torus to the unbound region. After it enters into the unbound region, it ionizes (solely due to the action of the field). A nonsequential double ionization corresponds to the situation when the exchange of energy allows the inner electron to jump from the bound region to the unbound one, while the outer electron keeps enough energy to remain ionized.  The amount of energy exchanged as well as the size of the bound region are the two main reasons for the sensitivity of the nonsequential double ionization to changes in the parameters $\alpha$ and $\delta$.

Using an equal sharing of the energy of the inner and outer electron~\cite{Maug09}, we derive a prediction for the intensity when the CDI probability is maximum. At $(y,p_y)=(0,0)$, when the inner electron is bounded, its energy is $U_{{\rm in}} = -2$, while at $(y,p_y)=(y_m,0)$, when the inner electron is on the boundary  (on the invariant torus with action $A_m$), its energy is $U_{{\rm in}}= -2/\sqrt{y_{m}^{2}+1}$. This leads to a difference in energy of $\Delta {\cal E}_y = 2 - 2/\sqrt{y_m^2+1}$ which is the minimum energy required to ionize the inner electron in one recollision. This energy comes from the outer electron, and when
\begin{equation} \label{eq:equalSharing}
   \Delta {\cal E}_y=\frac{{\cal E}_{\rm max}}{2}, 
\end{equation}
the outer electron gives enough energy to ionize the inner electron (in one recollision) while keeping itself ionized as well.

\begin{figure}
   \centering
   \includegraphics[width = \linewidth]{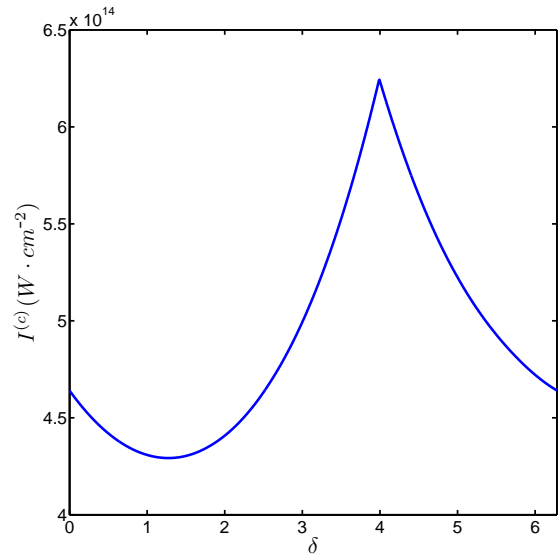}
   \caption{ \label{fig:E_expansion} 
   Expansion of $I^{\rm (c)}$, where $\kappa$ and $M$ have been solved for numerically. For this plot $\alpha=0.45$, $k=3$, and $\omega=0.0584$.}
\end{figure}

The equal sharing relation can be solved in a manner very similar to the case of the one laser system ~\cite{Maug09} by making the assumption that double ionization occurs at peak field amplitude, as given by Eq.~(\ref{eq:expansionApprox}). 
Combining Eq.~(\ref{eq:equalSharing}) with Eq.~(\ref{eq:expansionApprox}) yields the following approximation for $E_0^{\left(c\right)}$
\begin{equation} \label{eq:EqualSharingRelation}
   E_{0}^{\rm (c)} = \frac{4\omega}{\sqrt\kappa} - {\left(\frac{2\omega}{\sqrt\kappa}\right)}^\frac{3}{2} \sqrt{M} + \mathcal{O}\left( \left( \frac{2 \omega}{\sqrt\kappa} \right) ^ 3 \right).
\end{equation}
For numerically computed $\kappa$ and $M$ we can see from Fig.~\ref{fig:E_expansion} that $I^{\rm (c)}$ does not vary significantly with $\delta$. Over the entire range of $\delta$, the critical intensity $I^{\rm (c)}$ varies by approximately $2\times 10^{14}\ {\rm W} \cdot {\rm cm}^{-2}$ which is compatible with the variations of $I^{\rm (c)}$ as observed in Fig.~\ref{fig:knee_delta}. This variation is small compared to the variations of the laser intensity from $10^{13}$ to $10^{17}\ {\rm W} \cdot {\rm cm}^{-2}$ over the entire CDI and UDI regimes. It is also small compared to the variations of $I^{(t)}$ which are of the order of $10^{16}\ {\rm W} \cdot {\rm cm}^{-2}$. It confirms Observation~\ref{Obs:2} from the statistical analysis, in Sec.~\ref{sec:stat}, namely that the knee regime (and particularly the intensity for which CDI is maximum) is fairly insensitive to the laser parameters.

\section{Conclusion}

Using the statistical and nonlinear dynamical analysis we showed that the laser parameters $\alpha$ and $\delta$ can be used to control both the probabilities of double ionization and the rates of change of these probabilities both in the CDI and UDI regimes. This control acts on the outer electron by regulating the amount of energy it can bring to the core during a recollision. For the inner electron, the control depends on its stability, by changing the size of the bound area. Those two modulations act together to determine the double ionization probability, but it is the inner electron which has the strongest impact on its variations. In a nutshell, using the inner and outer electrons reduced models, we were able to explain the three observations obtained with the statistical analysis of the system: Namely the high sensitivity of the double ionization yield and saturation of double ionization in the choice of the laser parameters and the low sensitivity of the knee regime in the same parameters. This control helps in regulating the proportion of sequential versus nonsequential processes which occur in this atomic system.

\section{Acknowledgements}
C.C. and F.M. acknowledge financial support from the PICS program of the CNRS. This work is partially funded by NSF.




\end{document}